# The fine structure of the vortex-beams in the biaxial and biaxially-induced birefringent media caused by the conical diffraction

T. Fadeyeva, C. Alexeyev, P. Anischenko and A. Volyar

*Department of Physics, Taurida National V.I. Vernadsky University*
*Vernadsky av.4, Simferopol, Ukraine, 95007*
*e-mail: volyar@crimea.edu*

## Abstract

We consider the paraxial propagation of nondiffracting singular beams inside natural biaxial and biaxially-induced birefringent media in vicinity of one of the optical axes in terms of eigenmode vortex-beams, whose angular momentum does not change upon propagation. We have predicted a series of new optical effects in the natural biaxial crystals such as the stable propagation of vector singular beams bearing the coupled optical vortices with fractional topological charges, the conversion of the zero-order Bessel beam with a uniformly distributed linear polarization into the radially-, azimuthally- and spirally-polarized beams and the conversion of the space-variant linear polarization in the combined beam with coupled vortices. We have revealed that the field structure of the vortex-beams in the biaxially-induced crystals resembles that in the natural biaxial crystals and form the vector structure inherent in the conical diffraction. However, the mode beams in this case do not change the propagation direction as it takes place inside natural biaxial crystals.

## I. Introduction

Theoretically predicted by W. R. Hamilton as early as in 1832 [1] and experimentally revealed by H. Lloyd in the same year [2], the internal conical refraction remains up to now one of the brightest optical phenomenon of modern optics [3]. The classical manifestation of the internal conical refraction is that the unpolarized bundle of rays spreading parallel to one of the optical axes of a biaxial crystal is transformed into the hollow conical beam inside the crystal and emerges from the crystal in the form of the tube beam [4]. In terms of geometrical representation, the conical refraction is controlled by four points at the positions of intersections of two wave surfaces [4] located on two optical axes. The intersection vicinity takes the form of a double cone. The intersections themselves are conventionally called the diabolic points (see, e.g. [10] and references therein). These organize the centers of degeneracy where a great number of rays (directions of the Poynting vectors) propagating along the conical surface corresponds to one wave vector of the plane wave at the crystal input face. Naturally, such a ray uncertainty could not but affect the topological structure of the beam wave spreading along the crystal.

The vector beam representation of the phenomenon was at first described in the papers [5,6] by Belsky and Hapalyuck for the linearly polarized plane wave truncated by the circular diaphragm and the Hermite-Gaussian beams, respectively, on the base of the spectral integral approach. This representation was theoretically and experimentally extended over the ring structure of the conical refraction of the Gaussian beam in the papers [5-7]. Berry et al. [9-11] analyzing the propagation of the paraxial Gaussian beam along the crystal optical axis have brought to light a fine structure of the Pogendorff double-ring with central intensity maximum shaped by the far diffraction field outside the crystal. In the papers [8] the authors using the spectral integral method have shown that the structural transformations in nondiffracting beams inside the crystal are caused by the energy exchange between the circularly polarized beam components bearing the optical vortices that, in fact, enable us to shape the singly charged vortex-beam using the initial Gaussian beam

However, similar structural transformations of the wave beams are not exclusively restricted to the natural biaxial crystals. Vlokh et al. [12] have recently shown that the uniaxial crystal twisted around its optical axis manifests the features of the biaxial crystal forming much the same patterns of the space variant polarization. The mechanical twist of the uniaxial crystal induces the biaxial-like state in the initial uniaxial crystal with a complex pattern of the local axes orientation. We will call them later on the *biaxially-induced crystals.*

Appearance of the optical vortex in one of the beam components while the initial circularly polarized beam has no optical vortex is associated with appearance of the orbital angular momentum in the output beam [13]. Berry et al. [14] have shown that the initial Gaussian beam in the result of the conical diffraction acquires the fractional orbital angular momentum per photon equal to $1/2\hbar$ at the crystal output. The corresponding experiments with the fractional angular momentum of the beam caused by the conical diffraction were demonstrated in the paper [15]. However, similar transformation can be also caused in the beam components bearing the optical vortices with the topological charges of $l = \pm 1/2$. In favor of existence of such a beam state says a simple geometrical construction similar to that in uniaxial crystals for the radially and azimuthally polarized beams [19]. Indeed, there are two eigen linear polarizations of the plane waves at each point of the radiation field inside the biaxial crystal. For the case of the conical diffraction the eigen polarizations are parallel and perpendicular to the plane formed by the optical axis and the ray passing through the given point [16]. The patterns of the beam fields of the conical diffraction constructed in the above described way represent the non-uniformly polarized fields with a linear polarization in each point so that the topological index of the central polarization singularity can be $s = \pm 1$. The polarization distributions of the beam fields obtained in the papers [5-9] are of the non-uniformly polarized pattern with different elliptical polarization states at each point while the non-uniform pattern with a linear polarization in each point cannot be shaped by such mode beams in principle. At the same time, the space variant linear polarization distribution is at sight formed by the beam components bearing the coupled vortices with half-integer indices (see Sec.2.4). At the first glance it seems that using the mode beams with the half-integer vortex topological charge lacks the physical sense because such wave fields are not stable when propagating. A lot of papers show theoretically and experimentally (see e.g. [17,18]) that the wave field with such a scalar vortex state splits into a great number of singular beams bearing standard vortices when transmitting through isotropic homogeneous medium or free space. But it does not mean at all that half integer vortices cannot exist in a coupled state in the vector singular beams steadily propagating through anisotropic medium. When breaking the coupled state down, the beam comes to the unstable one forming new steady states.

All above discussed problems point out potential possibilities of the conical refraction phenomenon in shaping singular beams with fine topological structure and unusual properties. In order to bring to light these hidden field structures, for example, the spirally polarized beams with continuously tunable structure, it is necessary probably to use appropriate physical approaches different from the standard method of the spectral integral.

The aim of this paper is twofold: 1) to consider the conversion processes of the paraxial nondiffracting beams in terms of the eigen mode beams both inside the biaxial and biaxially-induced birefringent media exposing their generalities and differences and 2) to analyze characteristic features of the stable propagation of the nondiffracting vector beam states with the coupled optical vortices with fractional topological charges in the field components.

## II. Natural biaxial birefringent medium

*2.1 The basic paraxial wave equation*

We will consider at first the propagation of monochromatic light beams at the frequency $\omega$ through an unbounded biaxial birefringent medium with the permittivity tensor written in the major crystallographic axes $(x_1, y_1, z_1)$ (see Fig.1) in the diagonal form. We focus our attention on the nondiffracting beams with the electric field: $\mathbf{E}(x, y, z) = \mathbf{E}(x, y)\exp(-i\beta z)$ transmitting inside the medium along the z-axis at the angle $\theta$ to the $z_1$-axis of the crystal coordinate system as it is shown in Fig.1. We restrict ourselves to the paraxial case when the longitudinal component $E_z(x, y)$ is very small in comparison with the transverse $\mathbf{E}_\perp(x, y)$ components.

In the $(x, y, z)$ coordinates the permittivity tensor takes the form:

$$\hat{\varepsilon} = \begin{pmatrix} \varepsilon_{11}\cos^2\theta + \varepsilon_{33}\sin^2\theta & 0 & (\varepsilon_{11}-\varepsilon_{33})\sin\theta\cos\theta \\ 0 & \varepsilon_{22} & 0 \\ (\varepsilon_{11}-\varepsilon_{33})\sin\theta\cos\theta & 0 & \varepsilon_{11}\sin^2\theta + \varepsilon_{33}\cos^2\theta \end{pmatrix} = \begin{pmatrix} \varepsilon_1 & 0 & -\varepsilon_{13} \\ 0 & \varepsilon_2 & 0 \\ -\varepsilon_{13} & 0 & \varepsilon_3 \end{pmatrix}. \quad (1)$$

where we assume that $\varepsilon_{33} > \varepsilon_{22} > \varepsilon_{11}$. The condition $div\,\mathbf{D} = div(\hat{\varepsilon}\mathbf{E}) = 0$ can be written for the nondiffracting beam in the form

$$\partial_x(\varepsilon_1 E_x - \varepsilon_{13} E_z) + \varepsilon_2 \partial_y E_y - i\beta(\varepsilon_3 E_z - \varepsilon_{13} E_x) = 0. \quad (2)$$

The terms in eq. (2) with the factor $\beta$ are much greater than the rest terms in the equation in the paraxial approximation and we can reduce the above equation to

$$E_z \approx \frac{\varepsilon_{13}}{\varepsilon_3} E_x. \quad (3)$$

In the above approximation, we must bear in mind that the parameter $\varepsilon_{13}$ cannot be very small because in eq. (3) we neglect the term in the form $(\varepsilon_2 - \varepsilon_1)\nabla_\perp \mathbf{E}_\perp$. When decreasing the parameter $\varepsilon_{13}$ the angle between the optical axes decreases too. But the term $(\varepsilon_2 - \varepsilon_1)\nabla_\perp \mathbf{E}_\perp$ is responsible for the generation of the doubly charged vortex-beams in uniaxial crystals [19]. It means that the angle range in vicinity of $\theta = 0, \pi/2$ is out of our analysis so that we will not take into account the competitive effects in the processes of the vortex generation. Besides, we rewrite the factor $\nabla \mathbf{E}$ as $\nabla \mathbf{E} = \nabla_\perp \mathbf{E}_\perp - i\beta E_z$. Thus, in the paraxial approximation for the nondiffracting beams the wave equation

$$\nabla^2 \mathbf{E} + k^2 \mathbf{D} = \nabla(\nabla \mathbf{E}) \quad (4)$$

with $k = \omega/c$, $c$ is the speed of light, can be written for the transverse components as

$$\nabla^2 E_x + k^2 \bar{\varepsilon}\, E_x = \gamma E_x - i\alpha \partial_x E_x, \quad (5)$$

$$\nabla^2 E_y + k^2 \bar{\varepsilon}\, E_y = -\gamma E_y - i\alpha \partial_y E_x, \quad (6)$$

where

$$\gamma = k^2 \Delta \varepsilon, \quad \alpha = \beta \frac{\varepsilon_{13}}{\varepsilon_3}, \qquad (7)$$

$$\Delta \varepsilon = \frac{\varepsilon_2 - \varepsilon_1 + \varepsilon_{13}^2 / \varepsilon_3}{2}, \quad \overline{\varepsilon} = \frac{\varepsilon_1 - \varepsilon_{13}^2 / \varepsilon_3 + \varepsilon_2}{2}. \qquad (8)$$

Our interest is the circularly polarized beam components bearing the optical vortices i.e. the field components in the form

$$E_+(r,\varphi) = F_+(r)\exp(i\,p_+\varphi), \; E_-(r,\varphi) = F_-(r)\exp(i\,p_-\varphi), \qquad (9)$$

where $F_+, F_-$ are the wave amplitudes, $p_+, p_-$ stand for the vortex topological charges and

$$E_+ = E_x - iE_y, \; E_- = E_x + iE_y. \qquad (10)$$

It is convenient for our analysis to choose new coordinates in the form [19]:

$$u = x + i\,y, \; v = x - i\,y \qquad (11)$$

so that

$$2\partial_u = \partial_x - i\partial_y, \; 2\partial_v = \partial_x + i\partial_y, \; \nabla^2 \equiv 4\partial_{uv}^2 + \partial_{zz}^2. \qquad (12)$$

Then we obtain the paraxial wave equations in the form

$$\{4\partial_{uv}^2 + U^2\}E_+ = \gamma E_- - i\alpha\left(\partial_u E_+ + \partial_u E_-\right), \qquad (13)$$

$$\{4\partial_{uv}^2 + U^2\}E_- = \gamma E_+ - i\alpha\left(\partial_v E_+ + \partial_v E_-\right) \qquad (14)$$

with $U^2 = k^2 \overline{\varepsilon} - \beta^2$.

*2.2 Eigen vortex-beams in the conical diffraction process*

However, the equation obtained cannot answer the question how the vortex-beams (9) are transmitted through the crystal? To understand it let us consider the actions implemented by the operators $\partial_u$, $\partial_v$ and $\partial_{uv}^2$ in eqs (13) and (14).

The operators $\partial_u$ and $\partial_v$ are written in the polar coordinates $(r,\varphi)$ as

$$\partial_u \equiv \frac{e^{-i\varphi}}{2}\left(\partial_r - \frac{i}{r}\partial_\varphi\right), \; \partial_v \equiv \frac{e^{i\varphi}}{2}\left(\partial_r + \frac{i}{r}\partial_\varphi\right). \qquad (15)$$

Thus, the operator $\partial_u$ decreases the topological charge of the vortex beam by a unity while the operator $\partial_v$ increases the vortex topological charge by a unity in the right sides of the equations (13) and (14). The operator $\partial_{uv}^2$ in the left sides of eqs (13) and (14) leaves the vortex uncharged. At the same time, the first terms in the right side of eqs (13) and (14) with the components $E_-$ and $E_+$ are not subjected to acting the operators $\partial_u$ and $\partial_v$. The processes of the vortex birth and death are not compensated in the right sides of the equations and, consequently, the equations (13) and (14) cannot describe evolving the vortex-beams in the form (9).

In order to avoid the contradiction in the right sides of the eqs (13) and (14) we assume that

$$\gamma E_- = i\alpha \partial_u E_+, \quad \gamma E_+ = i\alpha \partial_v E_-. \tag{16}$$

Then the equations (13) and (14) take on the form

$$\{4\partial^2_{uv} + U^2\} E_+ = -i\alpha \partial_u E_-, \tag{17}$$

$$\{4\partial^2_{uv} + U^2\} E_- = -i\alpha \partial_v E_+. \tag{18}$$

Substituting eq. (9) into eqs (17) and (18) with $p_+ = p$, $p_- = p+1$ we find the equations for amplitudes $F_\pm$ in the polar coordinates:

$$\left( \frac{d^2}{dr^2} + \frac{1}{r}\frac{d}{dr} - \frac{p^2}{r^2} + U^2 \right) F_+(r) = -i\frac{\alpha}{2}\left[ \frac{dF_-(r)}{dr} + \frac{p+1}{r} F_-(r) \right], \tag{19}$$

$$\left( \frac{d^2}{dr^2} + \frac{1}{r}\frac{d}{dr} - \frac{(p+1)^2}{r^2} + U^2 \right) F_-(r) = -i\frac{\alpha}{2}\left[ \frac{dF_+(r)}{dr} - \frac{p}{r} F_+(r) \right]. \tag{20}$$

We search for the solutions of the above equations in the form

$$F_+ = Z_p(Vr), \quad F_- = i\sigma_1 Z_{p+1}(Vr), \quad \sigma_1 = \pm 1, \tag{21}$$

where $Z_p(Vr)$ stands for the Bessel functions $J_p(Vr)$ and $Y_p(Vr)$ of the first and the second kind, respectively, $V$ is an unknown wave parameter.

The equation (19) and (20) can be rewritten now as

$$\left( \frac{d^2}{dr^2} + \frac{1}{r}\frac{d}{dr} - \frac{p^2}{r^2} + V^2 \right) Z_p(Vr) = 0, \tag{22}$$

$$\left( \frac{d^2}{dr^2} + \frac{1}{r}\frac{d}{dr} - \frac{(p+1)^2}{r^2} + V^2 \right) Z_{p+1}(Vr) = 0, \tag{23}$$

where

$$V^2 = U^2 - \sigma_1 \frac{\alpha}{2} V \tag{24}$$

and we made use of the recurrent relations

$$\frac{dZ_p(r)}{dr} + \frac{p}{r} Z_p(r) = Z_{p-1}(r), \quad \frac{dZ_p(r)}{dr} - \frac{p}{r} Z_p(r) = -Z_{p+1}(r). \tag{25}$$

From eq. (24) we find

$$V = -\sigma_1 \frac{\alpha}{4} \pm \sqrt{\left(\frac{\alpha}{4}\right)^2 + U^2}. \tag{26}$$

On the other hand, the conditions (16) can be rewritten in the form of two equations

$$\partial^2_{uv} E_\pm + \frac{\gamma^2}{\alpha^2} E_\pm = 0. \qquad (27)$$

Taking into account eqs (9) we obtain

$$\left( \frac{d^2}{dr^2} + \frac{1}{r}\frac{d}{dr} - \frac{p^2}{r^2} + 4\frac{\gamma^2}{\alpha^2} \right) F_+ = 0, \qquad (28)$$

$$\left( \frac{d^2}{dr^2} + \frac{1}{r}\frac{d}{dr} - \frac{(p+1)^2}{r^2} + 4\frac{\gamma^2}{\alpha^2} \right) F_- = 0. \qquad (29)$$

Eqs (22) and (23) are consistent with eqs (28) and (29) provided that

$$V^2(\beta) = 4\frac{\gamma^2}{\alpha^2(\beta)} \quad \text{or} \quad V(\beta) = 2\sigma_2 \frac{\gamma}{\alpha(\beta)} \qquad (30)$$

from whence we come up at the equation for the propagation constant $\beta^2$:

$$\beta^2 = k^2 \left\{ \frac{\bar{\varepsilon}}{2} - \sigma_1 \sigma_2 \frac{\Delta \varepsilon}{2} + \sigma_3 \sqrt{\left(\frac{\bar{\varepsilon}}{2} - \sigma_1 \sigma_2 \frac{\Delta \varepsilon}{2}\right)^2 - 4\frac{\Delta \varepsilon^2}{\Delta^2}} \right\}, \quad \sigma_{2,3} = \pm 1. \qquad (31)$$

where we used eqs (7) and the designation $\Delta = \varepsilon_{13}/\varepsilon_3$.

The equation (31) can be regarded as the dispersion equation for the propagation constant $\beta^2$ as a function of the crystal parameters and the angle $\theta$ (see eq.(1)) of the beam propagation. Both the propagation constant $\beta^2 = \beta^2(\sigma_1, \sigma_2, \sigma_3)$ and the wave parameter $V = V(\sigma_1, \sigma_2, \sigma_3)$ depend on the signs of the $\sigma_{1,2,3}$. We can designate the vortex-beam states as $(\sigma_1, \sigma_2, \sigma_3)$ so that we have eight states of the vortex beam. However, it is easy to show that the states $(1,-1,1) = (-1,1,1)$; $(1,1,-1) = (1,-1,-1)$; $(1,1,-1) = (-1,-1,-1)$ are degenerated and we deal only with four beam states. *It is important to emphasis that the propagation constant $\beta$ and the wave parameter V do not depend on the topological charge p of the vortex-beam.*

The transverse field components can be now presented in the form

$$E_+(r) = \left[ C_1 J_p(Vr) + C_2 Y_p(Vr) \right] \exp(ip\varphi) \exp(-i\beta z), \qquad (32)$$

$$E_-(r) = i\sigma_1 \left[ C_1 J_{p+1}(Vr) + C_2 Y_{p+1}(Vr) \right] \exp[i(p+1)\varphi] \exp(-i\beta z), \qquad (33)$$

$C_{1,2}$ are constants.

### 2.3 Beams with the integer vortex topological charge

We assume that $p = n$ is the integer. In this case the Bessel function of the second order $Y_m(Vr) \xrightarrow{r \to 0} \infty$ and we set $C_2 = 0$, $C_1 = 1$ so that the field equations come to the form

$$\mathbf{E} = \begin{pmatrix} E_+ \\ E_- \end{pmatrix} = \begin{pmatrix} J_m[V(\sigma_1 \cdot \sigma_2, \sigma_3)r] \\ i\sigma_1 J_{m+1}[V(\sigma_1 \cdot \sigma_2, \sigma_3)r]\exp[i\varphi] \end{pmatrix} \exp(im\varphi)\exp[-i\beta(\sigma_1 \cdot \sigma_2, \sigma_3)z]. \quad (34)$$

The field structure in eq. (34) coincides in general features with that obtained in the paper [6]. The right hand polarized component (RHP) carries over the optical vortex with the topological charge $p_+ = m$ whereas the left hand polarized component has the vortex with $p_- = m+1$. The difference between the results is in the propagation constants $\beta$ and the wave parameters $V$. The fact is that the authors of the paper [6] consider only the case with the Bessel beam before the crystal propagating along the optical axis focusing no attention on the beam parameters and the dispersive relation. We analyze here different variations of the Bessel beams transmitting at different angles $\delta - \theta$ (see Fig.1) to the optical axis inside the crystal that correspond to a lot of possibilities in the initial beam states before the crystal.

Typical dispersion curves $\beta(\theta)$ and $V(\theta)$ for eqs (30) and (31) are shown in Fig.2. As the biaxial birefringent medium we choose the aragonite crystal often used in the experiments (see e.g. [10]) with parameters $\sqrt{\varepsilon_{11}} = n_1 = 1.533$, $\sqrt{\varepsilon_{22}} = n_2 = 1.686$, $\sqrt{\varepsilon_{33}} = n_3 = 1.691$. The curves $\beta(\theta)$ form two closed circuits inside the angle range $\Delta\theta$. In the framework of our approach we cannot say what beam states can exist outside the angle range $\Delta\theta$, but they are not the vortex-beam modes described by eq. (9). The physical base of it is rather evident – the birefringent crystal cannot preserve circularly polarized beam components far from optical axes propagating with the same phase velocities (see e.g. [20]). The dispersion curves $\beta(\theta)$ in the states $(-1,1,1)$ and $(1,1,1)$ cross themselves at the angle $\theta$ corresponding to the angle position of optical axis $\theta = \delta$ (see Fig.1):

$$tg^2\delta = \frac{\varepsilon_{33}(\varepsilon_{22} - \varepsilon_{11})}{\varepsilon_{11}(\varepsilon_{33} - \varepsilon_{22})} \quad (35)$$

so that the propagation constants of these states are the same: $\beta(1,1,1) = \beta(-1,1,1)$. At the same time, the dispersion curves $V(\theta)$ in the above states only touch each other at this angle position and the wave parameters are much the same in the broad angle range near optical axis.

It is noticeable that the curves $\beta(\theta)$ in the states $(-1,-1,-1)$ and $(-1,1,-1)$ correspond to the zero propagation constants $\beta(-1,-1,-1) = \beta(-1,1,-1) = 0$ at the angle $\theta = \delta$. This means that the propagating waves vanish and replaced by evanescent waves. At the edges of the angle range $\Delta\theta$, the dispersion curves get connected to each other for all states. However, at the framework of our approach we cannot expand our quantitative consideration over the states

$(-1,-1,-1)$ and $(-1,1,-1)$. The fact is that we restrict ourselves to the paraxial approximation where the longitudinal $E_z$ component is very small. As the propagation constant $\beta(\theta)$ decreases the contribution of the $E_z$ component increases. At the zero propagation constant, the $E_z$ component plays the key part. It makes us to limit ourselves to only quantitative remarks.

The characteristic features of the polarization distribution in the mode vortex-beams for different wave parameters are illustrated by Fig.3. Near the beam axis, it is observed predominance of the circular polarization that takes turns into elliptic polarizations with different directions of the major ellipse axis shaping typical pattern of the polarization singularity in the form of lemon [37] with topological index ½. Far from the beam axis the smooth polarization distribution is periodically modulated by the ring dislocations of the nondiffracting beam. Consider several cases of the complex vortex-beams.

*Case 1. The uniformly circular polarized vortex-beam.* Using the states $(1,1,1)$ and $(-1,1,1)$ we can describe the evolution of the vortex beam with the initial circular polarization. Note that the wave parameters $V$ are much the same $V(1,1,1) \approx V(-1,1,1)$ (see Fig.2) in the paraxial region while the propagation constants are essentially different $\beta_+ - \beta_- = \Delta\beta$. By superposing these states at the initial plane $z = 0$ we obtain the evolution process of the combined vortex beam along the $z$ axis:

$$\mathbf{E} = \mathbf{E}(1,1,1) + \mathbf{E}(-1,1,1) = \frac{e^{-i\bar{\beta}z}}{2} \left\{ \begin{pmatrix} J_m(Vr) \\ 0 \end{pmatrix} e^{im\varphi} \cos(\Delta\beta z) - \begin{pmatrix} 0 \\ J_{m+1}(Vr) \end{pmatrix} e^{i(m+1)\varphi} \sin(\Delta\beta z) \right\}, \quad (36)$$

where $\bar{\beta} = (\beta_+ + \beta_-)/2$. The state with the RHP and the *m*-charged vortex turns into the state with the LHP and (m+1)-charged vortex at the half beating length similar to that described at first in the papers [8].

*Case 2. The radially (TM modes) azimuthally (TE modes) and spirally polarized vortex-beams.* The TM and TE vortex-beams are of the eigen modes of the uniaxial crystal for the case of light propagation along the crystal optical axis [19]. However, in the biaxial crystal, such mode beams experience structural transformations. Consider them in details.

*TM-mode.* The wave structure of the TM mode in the *z=0* plane is $\mathbf{E}_{TE}(z=0) = (\mathbf{e}_+ e^{-i\varphi} + \mathbf{e}_- e^{i\varphi}) J_1(Vr)$. Assuming at first that the index $m = -1$ in eq. (36) we find

$$\mathbf{E}_1 = \frac{e^{-i\bar{\beta}z}}{2} \left\{ \begin{pmatrix} -J_1(Vr)e^{-i\varphi} \\ 0 \end{pmatrix} \cos(\Delta\beta z) - \begin{pmatrix} 0 \\ J_0(Vr) \end{pmatrix} \sin(\Delta\beta z) \right\}. \quad (36a)$$

Assuming now that $m = 0$ and after a little reconstruction of eq. (36), we obtain

$$\mathbf{E}_2 = \mathbf{E}_2(1,1,1) - \mathbf{E}_2(-1,1,1) = i\frac{e^{-i\bar{\beta}z}}{2}\left\{\begin{pmatrix}J_0(Vr)\\0\end{pmatrix}\sin(\Delta\beta z) + \begin{pmatrix}0\\J_1(Vr)e^{i\varphi}\end{pmatrix}\cos(\Delta\beta z)\right\}. \quad (36b)$$

Combining eqs (36a) and (36b) we come to the relation

$$\mathbf{E}_{TM} = \frac{e^{-i\bar{\beta}z}}{2}\left\{\begin{pmatrix}1\\1\end{pmatrix}J_0(Vr)\sin(\Delta\beta z) + \begin{pmatrix}e^{-i\varphi}\\e^{i\varphi}\end{pmatrix}J_1(Vr)\cos(\Delta\beta z)\right\}. \quad (36c)$$

*TE-mode.* The wave structure of the TE mode in the $z=0$ plane is $\mathbf{E}_{TE}(z=0) = (\mathbf{e}_+ e^{-i\varphi} - \mathbf{e}_- e^{i\varphi})J_1(Vr)$. Using the same approach we obtain

$$\mathbf{E}_{TE} = \frac{e^{-i\bar{\beta}z}}{2}\left\{-\begin{pmatrix}1\\1\end{pmatrix}J_0(Vr)\sin(\Delta\beta z) + \begin{pmatrix}e^{-i\varphi}\\-e^{i\varphi}\end{pmatrix}J_1(Vr)\cos(\Delta\beta z)\right\}. \quad (36d)$$

At the plane $z=0$ we have the unperturbed TM mode in eq. (36c) with the topological index $s=1$ of the polarization singularity at the beam axis. A little displacement from the $z=0$ plane results in the birth event of two lemons with topological indices $s=1/2$ (see Fig.4a). Besides, near the first dislocation ring we observe the birth event of two stars with the $s=-1/2$ indices (we focus our attention only on the polarization singularities near the beam axis). When propagating the beam, the lemons and stars draw together and annihilate to shape the zero-order Bessel beam with uniform linear polarization along the *y*-axis at the quarter beating length $\Lambda/4$. Then the uniformly polarized state is destroyed forming again the field with two lemons and two stars. At the half beating length $\Lambda/2$ lemons and stars annihilate again recovering the TM mode structure. The topological reactions in the TE mode run similarly with that difference that the uniform linear polarization at $z=\Lambda/4$ is directed along the $x-$axis.

It is worth noting that simple coordinate transformations show that the uniformly linear polarized field of the zero-order Bessel beam can be converted either into the TM or TE modes. Such a process can be useful for practical engineering of the radially and azimuthally polarized beam that finds a bleeding-edge application for trapping and transportation of microparticles (see e.g. the paper [34] and references therein).

As an example let us consider the structural transformations in the zero-order Bessel beam with uniformly distributed linear polarization directed at the angle $\psi$ relative to the $y$-axis. The conversion process in such a beam field develops in accordance with the equation:

$$\mathbf{E}_{spiral} = \frac{e^{-i\bar{\beta}z}}{2}\left\{\begin{pmatrix}e^{i\psi}\\e^{-i\psi}\end{pmatrix}J_0(Vr)\cos(\Delta\beta z) + \begin{pmatrix}e^{-i(\varphi+\psi)}\\-e^{i(\varphi+\psi)}\end{pmatrix}J_1(Vr)\sin(\Delta\beta z)\right\} \quad (36e)$$

At the initial $z=0$ plane, the beam field has a uniform linear polarization directed at the angle $\psi$ to the $y-$axis (see Fig. 4b). Then the polarization singularities appear at different points of the beam cross-section distorting the uniform polarization distribution so that the spirally polarized pattern is formed at the quarter beating length. At the distances multiple to the half beating length, the field recovers its initial structure.

*2.4 Beams with the non-uniform linearly polarized distribution*

As was said in Introduction, the eigen plane waves of the conical diffraction have possibility to form the non-uniform polarization distribution with a linear polarization in each point. Such a field structure cannot be originated by the vortex-beams with the integer topological charges because the RHP and LHP components are described by different radial functions. However we can construct such a beam field at the z=0 plane with the help of the half-integer vortex beams in each field component

$$\mathbf{E} = \begin{pmatrix} \exp(i\varphi/2) \\ i\sigma_1 \exp[-i\varphi/2] \end{pmatrix} J_{1/2}\left[V(\sigma_1 \cdot \sigma_2, \sigma_3)r\right]. \tag{37}$$

that can be described in terms of the spherical Bessel function $j_0(x)$ [21]:

$$\mathbf{E}(z=0) = \sqrt{\frac{2Vr}{\pi}} \begin{pmatrix} \exp(i\varphi/2) \\ i\sigma_1 \exp[-i\varphi/2] \end{pmatrix} j_0(Vr), \tag{38}$$

where

$$j_0(x) = \sqrt{\frac{\pi}{2x}} J_{1/2}(x) = \sqrt{\frac{\pi}{2x}} Y_{-1/2}(x) = \frac{\sin x}{x}$$

are the spherical Bessel functions of the first kind. To trace the evolution of the beam (38) it is necessary to present such a combined wave field in terms of eigen modes (32) and (33). First of all, the solutions (32), (33) and the dispersion equation (31) can be applied not only for the integer indices $p = \pm m$, but also for the $p = \pm n/2$ indices with $n = 1, 3, 5...$. The beam state (38) at arbitrary plane $z = z_0$ can be written in the form of the eigen beams superposition:

$$\mathbf{E}(r,\varphi,z) = |1\rangle + |2\rangle = \frac{e^{-i\bar{\beta}z}}{2}\left\{\begin{pmatrix} e^{i\varphi/2} \\ e^{-i\varphi/2} \end{pmatrix} J_{1/2}(Vr)\cos(\Delta\beta z) + \begin{pmatrix} e^{-i3\varphi/2} \\ -e^{i3\varphi/2} \end{pmatrix} J_{3/2}(Vr)\sin(\Delta\beta z)\right\} \tag{39}$$

where

$$|1\rangle = \frac{e^{-i\bar{\beta}z}}{2}\left\{\begin{pmatrix} e^{i3\varphi/2} \\ 0 \end{pmatrix} J_{3/2}(Vr)\sin(\Delta\beta z) + \begin{pmatrix} 0 \\ e^{-i\varphi/2} \end{pmatrix} J_{1/2}(Vr)\cos(\Delta\beta z)\right\},$$

$$|2\rangle = \frac{e^{-i\bar{\beta}z}}{2}\left\{\begin{pmatrix} e^{i\varphi/2} \\ 0 \end{pmatrix} J_{1/2}(Vr)\cos(\Delta\beta z) + \begin{pmatrix} 0 \\ -e^{i3\varphi/2} \end{pmatrix} J_{3/2}(Vr)\sin(\Delta\beta z)\right\} \tag{40}$$

and we made use of the relation [21]: $Y_{\pm m+1/2}(x) = \mp(-1)^m J_{\mp m+1/2}(x)$.

Nevertheless, the question is still open: are the vortex beams with the $n/2$ topological charges stable when propagating?

The vortices with $n/2$-topological charges in the field components of the eigen modes

$$\mathbf{E} = \begin{pmatrix} \exp\left(i\frac{n}{2}\varphi\right) J_{n/2}(Vr) \\ i\sigma_1 \exp\left[i\left(\frac{n}{2}+1\right)\varphi\right] J_{n/2+1}(Vr) \end{pmatrix} e^{-i\beta z} \text{ or } \mathbf{E} = \begin{pmatrix} \exp\left(-i\frac{n}{2}\varphi\right) Y_{-n/2}(Vr) \\ i\sigma_1 \exp\left[i\left(-\frac{n}{2}+1\right)\varphi\right] Y_{-n/2+1}[Vr] \end{pmatrix} e^{-i\beta z} \quad (41)$$

are strongly coupled with each other by the biaxial birefringence of the medium. Certainly, they can be presented as an infinite superposition of the vortex beams (34) with the integer topological charges [17, 31] propagating at the defined angle $\theta$ corresponding to the defined values of the propagation constant $\beta$ and the wave parameter $V$. In contrast to homogeneous media and uniaxial crystals, the propagation constants $\beta$ and wave parameters $V$ do not depend on the value of the vortex topological charge in the considering case. All vortex-beams in the superposition propagate with the same phase velocity. The distortion of the field structure of the coupled vortex beam with $n/2$ - topological charge is absent. However, outside the biaxial crystal, the vortex-beams in the field components are decomposed into the infinite superposition of the integer-charged vortices running with different phase velocities. The state of $n/2$-charged vortex beam breaks down. Let us consider several the evolution of the linearly polarized combined beam (39).

The intensity and polarization distributions are shown in Fig.5. The polarization distribution illustrated by the figure has a linear polarization at each point of the field cross-section. The centered polarization singularity has the topological index $s = -1/2$. The pattern obtained reflects the conventional notion of the eigen liner polarizations of local plane waves propagating along different directions of the biaxial crystal [4, 16, 22]. However, such a space variant linear polarization cannot be preserved when propagating the beam as it was assumed early. The initial polarization structure is recovered at the beating length $\Lambda = 2\pi/\Delta\beta$. The corresponding structural evolution of the singular beam at the intermediate lengths is shown in Fig.6. A slight displacement from the z=0 plane results in appearing four pairs of the star -lemon with the topological indices $s = \pm 1$ in vicinity of each ring dislocation. As the beam propagates along the crystal, the lemons and stars are transposed and form the initial polarization distributions at the half beating length.

The vector field structure (37) consisting of two fractional vortex beams in the circularly polarized components has analogy in the Standard Model of particle physics, in particular, in the Gell-Mann's quark model of the hadrons [32]. Similar to the quark, the vortices of the fractional-integer order cannot exist outside the coupled states in the biaxial crystal forming another couple states when propagating in free space. However, unlike the quarks, the fractional-order vortex beam can be experimentally observed, in our opinion, at the output face of the crystal when exciting properly the input crystal face.

Naturally the stable propagation of the coupled vortex beams with half integer topological charges needs experimental verification. It should be noted here that the experimental creation of the scalar light beams bearing the half-integer vortices even at the given observation plane encounters a number of difficulties connected with structural instability of such states when propagating [31]. At the same time, those difficulties can be avoided when experimenting by using the vector wave beams shaped by the q-plates. The fact is that a typical distribution of the optical axes formed by the q-plate (see the paper [33] and references therein) has similar structure as that shown in Fig.6 for the linearly polarized distribution of the beam field (37). It could be useful for the experimental study of shaping the written above vector mode beams accompanying the conical diffraction. Let us consider one more useful case.

*Conversion of the vortex-beam with the initial linear polarization and the edge-like dislocations.*

Our aim is to trace the evolution of the field in the form at the z=0 plane:

$$\mathbf{E}(z=0) = \cos\frac{3\varphi}{2}\begin{pmatrix}1\\1\end{pmatrix} J_{3/2}(Vr). \tag{41a}$$

Such a linearly polarized singular beam with the edge-like dislocations can be composed in accordance with the following procedure. The beam with the state $|1\rangle = \mathbf{e}_+ J_{3/2}(Vr) e^{i\frac{3}{2}\varphi}$ at the z=0 plane can be presented as

$$|1\rangle = \frac{e^{-i\bar{\beta}z}}{2}\left\{\begin{pmatrix}e^{i3\varphi/2}J_{3/2}(Vr)\\0\end{pmatrix}\cos(\Delta\beta z) - \begin{pmatrix}0\\e^{i5\varphi/2}J_{5/2}(Vr)\end{pmatrix}\sin(\Delta\beta z)\right\}. \tag{41b}$$

Similar to that we can construct the state $|2\rangle$:

$$|2\rangle = \frac{e^{-i\bar{\beta}z}}{2}\left\{-\begin{pmatrix}e^{-i3\varphi/2}Y_{-3/2}(Vr)\\0\end{pmatrix}\cos(\Delta\beta z) - \begin{pmatrix}0\\e^{-i\varphi/2}Y_{-1/2}(Vr)\end{pmatrix}\sin(\Delta\beta z)\right\}. \tag{41c}$$

Two more states $|3\rangle$ and $|4\rangle$, we find as

$$|3\rangle = \frac{e^{-i\bar{\beta}z}}{2}\left\{\begin{pmatrix}0\\e^{i3\varphi/2}J_{3/2}(Vr)\end{pmatrix}\cos(\Delta\beta z) + \begin{pmatrix}e^{i\varphi/2}J_{1/2}(Vr)\\0\end{pmatrix}\sin(\Delta\beta z)\right\}, \tag{41d}$$

$$|4\rangle = \frac{e^{-i\bar{\beta}z}}{2}\left\{\begin{pmatrix}0\\e^{-i3\varphi/2}Y_{-3/2}(Vr)\end{pmatrix}\cos(\Delta\beta z) + \begin{pmatrix}e^{-i5\varphi/2}Y_{-5/2}(Vr)\\0\end{pmatrix}\sin(\Delta\beta z)\right\}. \tag{41e}$$

Then the required field structure takes on a form

$$\mathbf{E}(r,\varphi,z) = |1\rangle - |2\rangle + |3\rangle - |4\rangle = \frac{e^{-i\bar{\beta}z}}{2}\left\{\cos\frac{3\varphi}{2}J_{3/2}(Vr)\binom{1}{1}\cos(\Delta\beta z) + \right.$$
$$\left. + \binom{e^{i\varphi/2}J_{1/2}(Vr) - e^{-i5\varphi/2}J_{5/2}(Vr)}{e^{-i\varphi/2}J_{1/2}(Vr) - e^{i5\varphi/2}J_{5/2}(Vr)}\sin(\Delta\beta z)\right\}, \quad (41f)$$

The evolution of the intensity distribution of the light beam (41f) is illustrated in Fig.7. The lines of the zero amplitudes at the initial plane z=0 are directed along the rays $\varphi_1 = \pi/3$, $\varphi_2 = \pi$, $\varphi_3 = 5\pi/3$. A slight displacement from the initial plane results in appearing a number of discrete positions of the field with zero amplitude in vicinity of these rays. The uniformly linear polarized field is turned into the non-uniformly polarized one with the linear polarization at each point of the beam cross-section (see Fig.8). As the field propagates along the crystal the zero points experience slight displacements relative to their original positions while the linear polarization at each point remains unchanged. The interference patterns of the RHP and LHP field components near the beam axis shown in Fig.7 indicate a number of singly-charged optical vortices with opposite signs of the topological charges around the beam axis in each circularly polarized component, the opposite-charged vortices in the RHP and LHP components being placed at the same zero points. The half-integer optical vortices with opposite signs of the topological charges are positioned at the beam axis in the both field components. In the pattern of the polarization distribution shown in Fig.8 the degenerate polarization singularities with the topological indices $s = \pm 1$ positioned at the points of zero amplitudes gather around the central polarization singularity with $s = -1/2$. At the crystal lengths multiple of the half beating length $\Lambda$, the non-uniform polarization distribution vanishes replaced by the beam field with the uniformly distributed linear polarization.

### III.   Biaxially-induced birefringent medium

*3.1 The basic equation*

The property to generate the singly-charged optical vortices in one of the circularly polarized component while the other one has no optical vortices is not limited with biaxial crystals. As is well known the uniaxial crystals can generate the doubly-charged optical vortices in different types of paraxial beams [12,23,24] as the initial beam propagates along the optical axis. In order to generate the singly charged optical vortices in uniaxial crystals it is necessary to break down the symmetry of the system beam-crystal, for example, to use the linearly polarized beam tilted relatively to the optical axis [25]. The other approach was recently proposed by

Vlokh [7, 26-28] using the twisting of a uniaxial crystal around its optical axis. We will focus our attention here on the beam propagating through such a medium and trace the transformations of the major features of the vortex-beams in the biaxially-induced crystals in comparison with the natural biaxial crystals.

Let the circularly polarized (say, RHP) nondiffracting beam propagates along the optical axis of the uniaxial crystal while the crystal itself is subject to mechanical twisting around the optical axis (see Fig.9)). One end of the crystal is fastened while the second end is being twisted. The crystal twist caused by the torsion torque $M_z$ changes the piezo-optical coefficients of the medium and, consequently, the elements of the permittivity tensor $\hat{\varepsilon}$. As the result, the pattern of the polarization distribution inherent in the unperturbed uniaxial crystal [28] is drastically transformed. Instead of the centered doubly charged vortex, the singly charged vortex is shaped at the optical axis in the LHP beam component [12]. The polarization singularity of the radiation field with the topological index $s=1$ is replaced by that with the topological index $s=½$ (see Fig.9b). Such a polarization distribution is, generally speaking, inherent in the biaxial crystal. However, the authors of the paper [12] have shown that it is not the biaxial crystal in the traditional treatment. The situation proves to be more complex. The location of the optical axes of such a crystal is shown in Fig.10. The plane of the optical axes location is defined by the torsion axis and the ray passing through the given point. Moreover, the angle between the optical axes increases while growing the distance from the twisting axis. In fact we deal with the biaxially-like medium or, in other words, it is the biaxially-induced birefringent crystal with non-uniformly distributed local crystal axes.

Our aim is to describe the structure of the nondiffracting vortex-beams in such a biaxially-induced crystal in the paraxial case.

We start from the form of the permittivity tensor [29]:

$$\hat{\varepsilon} = \begin{pmatrix} \varepsilon_{11} - \pi_{14}\varepsilon_{11}^2\sigma_{32} & -\pi_{14}\varepsilon_{11}^2\sigma_{31} & -\pi_{14}\varepsilon_{11}\varepsilon_{33}\sigma_{31} \\ -\pi_{14}\varepsilon_{11}^2\sigma_{31} & \varepsilon_{11} + \pi_{14}\varepsilon_{11}^2\sigma_{32} & -\pi_{44}\varepsilon_{11}\varepsilon_{33}\sigma_{32} \\ -\pi_{14}\varepsilon_{11}\varepsilon_{33}\sigma_{31} & -\pi_{44}\varepsilon_{11}\varepsilon_{33}\sigma_{32} & \varepsilon_{33} \end{pmatrix}, \qquad (1)$$

where $\varepsilon_{11}, \varepsilon_{33}$ are the major dielectric permeabilities of the unperturbed uniaxial crystal, $\pi_{14}, \pi_{44}$ are the piezo-optic coefficients, $\sigma_{32} = \dfrac{2M_z}{\pi\rho^4}x$, $\sigma_{31} = \dfrac{2M_z}{\pi\rho^4}y$, $\rho$ stands for the crystal radius (see Fig.10).

As well as in the Section 2, we will restrict ourselves to the paraxial approximation when the longitudinal component $E_z$ of the field is very small. Then the wave equation (2.4) can be written for the transverse components in the form

$$\nabla^2 E_x + k^2 (\varepsilon - \gamma x) E_x - k^2 \gamma y\, E_y = 0, \qquad (2)$$

$$\nabla^2 E_y + k^2 (\varepsilon + \gamma x) E_y - k^2 \gamma y\, E_x = 0, \qquad (3)$$

where $\gamma = \pi_{14} \varepsilon^2 \dfrac{2M_z}{\pi \rho^4}$, $\varepsilon_{11} = \varepsilon$. In eqs (2) and (3) we neglected the right part $\nabla(\nabla \mathbf{E})$.

In the circularly polarized basis (2.10) we come to the equations

$$\nabla^2 E_+ + k^2 \varepsilon\, E_+ - k^2 \gamma r\, e^{-i\varphi}\, E_- = 0, \qquad (4)$$

$$\nabla^2 E_- + k^2 \varepsilon\, E_- - k^2 \gamma r\, e^{i\varphi}\, E_+ = 0, \qquad (5)$$

where $(r, \varphi)$ are the polar coordinates.

### 3.2 The beams with the half-integer topological charge

We will find at first the solutions for nondiffracting beams in the form

$$E_+ = \Psi_+(r) \exp\!\left[i\!\left(m_1 + \frac{k_1}{2}\right)\!\varphi\right] e^{-i\beta z},\ E_- = \sigma\, \Psi_-(r) \exp\!\left[i\!\left(m_2 + \frac{k_2}{2}\right)\!\varphi\right] e^{-i\beta z},\quad (6)$$

where $\sigma = \pm 1$.

Our interest here is in the solutions with the radial functions $\Psi = \Psi_+ = \Psi_-$ when eqs (4) and (5) can be reduced to the one scalar equation. It is possible only if

$$m_1 = -\frac{1 + k_1}{2},\ m_2 = \frac{k_2 - 1}{2},\ k_1 = k_2 = 1 \qquad (7)$$

so that the scalar equation is

$$\frac{d^2 \Psi}{dr^2} + \frac{1}{r}\frac{d\Psi}{dr} + \left(\bar{U}^2 + \sigma k^2 \gamma r - \frac{1}{4r^2}\right)\!\Psi = 0, \qquad (8)$$

$$\bar{U}^2 = k^2 \varepsilon - \beta^2, \qquad (9)$$

The real-valued solutions of eq. (3) are presented as

$$\Psi = \frac{C_1 Ai\!\left(-\dfrac{U^2 + \alpha R}{\alpha^{2/3}}\right) + C_2 Bi\!\left(-\dfrac{U^2 + \alpha R}{\alpha^{2/3}}\right)}{\sqrt{R}} \qquad (9)$$

in terms of the Airy functions $Ai(x)$ and $Bi(x)$ [21] with $U^2 = \rho_0^2 \bar{U}^2$, $\alpha = \sigma k^2 \gamma \rho_0^3$, $R = \dfrac{r}{\rho_0}$, $C_{1,2}$ being constants.

The Airy's functions can be presented as a superposition of Bessel functions of the order $\pm 1/3$. Near the axis $x \to 0$, the Airy functions can be described as

$$Ai(x) = a_1 f(x) - a_2 g(x),$$
$$Bi(x) = \sqrt{3}\left[a_1 f(x) + a_2 g(x)\right], \tag{10}$$

with

$$f(x) = 1 + \frac{1}{3!}x^3 + \frac{1\cdot 4}{6!}x^6 + \frac{1\cdot 4\cdot 7}{9!}x^9 + \ldots, \quad g(x) = x + \frac{2}{4!}x^4 + \frac{2\cdot 5}{7!}x^7 + \frac{2\cdot 5\cdot 8}{10!}x^{10} + \ldots,$$

$$a_1 = Ai(0) = Bi(0)/\sqrt{3} = 3^{-2/3}\Gamma(2/3) \approx 0.35550.$$

$$a_2 = Ai'(0) = Bi'(0)/\sqrt{3} = 3^{-1/3}\Gamma(1/3) \approx 0.25882.$$

The asymptotic values of the functions $x \to \infty$ are

$$Ai(-x) \sim \frac{1}{\pi^2}\frac{\sin\left(\frac{2}{3}x^{3/2} + \frac{\pi}{4}\right)}{x^{1/4}}, \quad Bi(-x) \sim \frac{1}{\pi^2}\frac{\cos\left(\frac{2}{3}x^{3/2} + \frac{\pi}{4}\right)}{x^{1/4}}, \quad |x| \gg 1 \tag{10}$$

so that there are not singularities of the function (10) $\Psi(r)$ in our case for $0 \leq r \leq \infty$, besides, $U^2 + \alpha R > 0$.

We require that $\Psi(r \to 0) \to 0$ in eq. (9), from whence we find

$$C_2 = 1, C_1 = C_0 = -Ai\left(-\frac{U^2}{\alpha^{2/3}}\right)/Bi\left(-\frac{U^2}{\alpha^{2/3}}\right). \tag{11}$$

The solution of the equations (4) and (5) is now

$$\mathbf{E} = \begin{pmatrix} \exp(-i\varphi/2) \\ \sigma\exp(i\varphi/2) \end{pmatrix} \frac{Ai\left(-\frac{U^2 + \alpha R}{\alpha^{2/3}}\right) - C_0 Bi\left(-\frac{U^2 + \alpha R}{\alpha^{2/3}}\right)}{\sqrt{R}}. \tag{12}$$

The polarization distribution at the $z=0$ plane in such a vortex-beam is described by the same pattern as that in the case of the vortex-beams in the biaxial crystals shown in Fig. 5. The field is linearly polarized at each point whereas the centered polarization singularity has the topological index $s=½$: the change of the azimuthal angle $\varphi$ by $\pi$ reduces the direction of the polarization state to $\pi/2$. However, the radial function modulating the radial field distribution of the nondiffracting beams is described now by the superposition of the Airy's functions shown in Fig.11 rather then the spherical Bessel function shown in Fig.5.

Unfortunately, *the vortex beams with the half-integer topological charges are not stable in the biaxially-induced crystals when propagating in contrast to those in the natural biaxial crystals*. The linearly polarized distribution pattern breaks rapidly down when propagating (see the next Section).

### 3.3 The beams with the integer topological charge

We will expand the search for the solutions of eqs (4) and (5) over the functions:

$$E_{\pm} = F_{+}(r)\exp\{im\varphi\}e^{-i\beta z}, \quad E_{\pm} = F_{+}(r)\exp\{i(m+1)\varphi\}e^{-i\beta z} \quad m = 0, \pm1, \pm2\ldots \quad (13)$$

Then the eqs (4) and (5) can be reduced to the form

$$\left\{\frac{d^2}{dr^2} + \frac{1}{r}\frac{d}{dr} - \frac{m^2}{r^2} + k^2\varepsilon\right\}F_{+}(r) + k^2\gamma\, rF_{-} = \beta^2 F_{+}, \quad (14)$$

$$\left\{\frac{d^2}{dr^2} + \frac{1}{r}\frac{d}{dr} - \frac{(m+1)^2}{r^2} + k^2\varepsilon\right\}F_{-}(r) + k^2\gamma\, rF_{+} = \beta^2 F_{-}. \quad (15)$$

To find the solution of the equations (14) and (15) we make use of the perturbation theory with degeneracy [30]. In the unperturbed (degenerated) case, we have $\gamma = 0$ so that the system (14), (15) can be presented as

$$\hat{H}_0 \begin{pmatrix} \tilde{F}_{+} \\ \tilde{F}_{-} \end{pmatrix} = \tilde{\beta}^2 \begin{pmatrix} \tilde{F}_{+} \\ \tilde{F}_{-} \end{pmatrix}, \quad (16)$$

where $\hat{H}_0 = \begin{pmatrix} \hat{h}_0 & 0 \\ 0 & \hat{h}_1 \end{pmatrix}$, $\hat{h}_0 = \frac{d^2}{dr^2} + \frac{1}{r}\frac{d}{dr} - \frac{m^2}{r^2} + k^2\varepsilon$, $\hat{h}_1 = \frac{d^2}{dr^2} + \frac{1}{r}\frac{d}{dr} - \frac{(m+1)^2}{r^2} + k^2\varepsilon$, $\tilde{F}_{\pm}$ and $\tilde{\beta}^2$ are the eigen functions and eigen values, respectively, of eqs (14), (15) with $\gamma = 0$. Evidently, there is a twofold degeneracy over the polarization states $|1\rangle = \begin{pmatrix} J_m(Ur) \\ 0 \end{pmatrix}$ and $|2\rangle = \begin{pmatrix} 0 \\ J_{m+1}(Ur) \end{pmatrix}$ (where $U$ is the wave parameter) with the same propagation constant $\tilde{\beta}^2$.

Generally speaking, the degenerated field state of the uniaxial crystal is [24] $|1\rangle = \begin{pmatrix} J_m(Ur) \\ 0 \end{pmatrix} e^{im\varphi}$, $|2\rangle = \begin{pmatrix} 0 \\ J_{m+2}(Ur) \end{pmatrix} e^{i(m+2)\varphi}$. However, when the longitudinal component of the initial beam $E_z$ is a very small, we can neglect the competition effects in the vortex beam between the uniaxial and biaxial crystal states (see also Section 2).

When "turning on" the perturbation the corresponding operator can be described as $\hat{V} = k^2\gamma\, r\, \hat{\sigma}_x$,

where $\hat{\sigma}_x = \begin{pmatrix} 0 & 1 \\ 1 & 0 \end{pmatrix}$ is the Pauli matrix. The full operator of the perturbed system is $\hat{H} = \hat{H}_0 + \hat{V}$, the solution of the perturbed system (14), (15) can be represented as

$$|F\rangle = C_1|1\rangle + C_2|2\rangle. \quad (17)$$

To find the correct form of the combination (17) it is necessary to average the full operator $\hat{H}$ over the basis $(|1\rangle, |2\rangle)$. We have for the $\hat{H}_0$ operator

$$\langle 1|\hat{H}_0|1\rangle = \langle 2|\hat{H}_0|2\rangle = \tilde{\beta}^2, \quad \langle 1|\hat{H}_0|2\rangle = \langle 2|\hat{H}_0|1\rangle = 0.$$

The operator of perturbation $\hat{V}$ has the following elements

$$\langle 1|\hat{V}|1\rangle = \langle 2|\hat{V}|2\rangle = 0,$$

$$\langle 1|\hat{V}|2\rangle = \langle 2|\hat{V}|1\rangle = C,$$

$$C = k^2 \gamma \frac{\iint_S r^2 J_m(Ur) J_{m+1}(Ur) dr}{\sqrt{\iint_S r J_m^2(Ur) dr \iint_S r J_{m+1}^2(Ur) dr}}, \tag{18}$$

and $S$ is the crystal cross-section.

Now the polarization correction $\Delta\beta^2$ to the propagation constant $\tilde{\beta}^2$ is defined by the characteristic equation

$$\hat{V}\mathbf{x} = \Delta\beta^2 \mathbf{x}, \text{ i.e. } C\hat{\sigma}_x \mathbf{x} = \Delta\beta^2 \mathbf{x}. \tag{19}$$

Its eigenvalues are

$$\Delta\beta_{1,2}^2 = \pm C \tag{20}$$

while the eigenvectors take the form

$$\mathbf{x}_1 = (1,1), \quad \mathbf{x}_2 = (1,-1). \tag{21}$$

From whence we have the correct form of the states

$$\begin{pmatrix} F_+ \\ F_- \end{pmatrix}_1 = \begin{pmatrix} J_m(Ur) \\ J_{m+1}(Ur) \end{pmatrix} \text{ for } \Delta\beta_1^2 \text{ and } \begin{pmatrix} F_+ \\ F_- \end{pmatrix}_2 = \begin{pmatrix} J_m(Ur) \\ -J_{m+1}(Ur) \end{pmatrix} \text{ for } \Delta\beta_2^2. \tag{22}$$

Thus, we find finally the form of the fields:

$$\begin{pmatrix} E_+ \\ E_- \end{pmatrix}_1 = \begin{pmatrix} J_m(Ur) \\ J_{m+1}(Ur)\exp(i\varphi) \end{pmatrix} \exp\{im\varphi\} \exp\{-i(\tilde{\beta} + \Delta\beta)z\} \tag{23}$$

and

$$\begin{pmatrix} E_+ \\ E_- \end{pmatrix}_1 = \begin{pmatrix} J_m(Ur) \\ -J_{m+1}(Ur)\exp(i\varphi) \end{pmatrix} \exp\{im\varphi\} \exp\{-i(\tilde{\beta} - \Delta\beta)z\}. \tag{24}$$

The dispersion curves (20) shown in Fig.12 are now defined the vortex topological charge $m$.

The obtained equations illustrate a typical structure of the paraxial nondiffracting beam propagating along the twisting axis of the twisted uniaxial crystal. These equations are completely identical to eqs (2.34) obtained in Section 2. This means that the observed phenomenon in the biaxially-induced crystal has much to do with the conical diffraction observed in the natural biaxial crystals. However, there is a radical difference. Indeed, the dispersion curves of the polarization correction $\delta\beta(U)$ shown in Fig.12 point out that the pattern of the diffraction is observed in the direction of the optical axis for either beam with arbitrary wave parameter $U$ while the conical refraction in the natural crystal is observed at the directions defined by the crystal depending on the parameter $V$ (see Fig.2).

Similar to the case of the natural biaxial crystal we can form the uniform polarization distribution bearing the *m*-charged optical vortex at the $z=0$ plane (see eq. (2.36)). The field structure is periodically transformed when propagating, recovering at the beating length $\Lambda = 2\pi / \Delta\beta$. In order to estimate the beating length $\Lambda$ we make use of the experimental data from the paper [12]. We suppose that the torsion torque $M_z = 0.162\, N \cdot m$ is applied to the *LiNbO$_3$* crystal with the ordinary refractive index $n_o = \sqrt{\varepsilon} = 2.3$ and piezo-optic coefficient $|\pi_{14}| = 8.87 \cdot 10^{-13}\, m^2 / N$. The crystal radius is $\rho = 0.5\, cm$. The polarization correction calculated from eqs (18) and (20) is $\beta = 20\, m^{-1}$ for $m = 1, U = 1000\, m^{-1}$. Thus, at the crystal length $L = \Lambda / 4 = 7.85\, cm$, the energy from the RHP vortex-beam component with the single topological charge is completely transformed into the LHP vortex-beam component with the double topological charge.

Recurring to the problem of stability of the half-integer vortex-beam it is worth remarking that polarization corrections to the propagation constant $\Delta\beta$ of standard vortices with the integer topological charges depend on the value of their topological charge (see Fig.12). As the result, ½-charged vortex-beam in each polarized component in eqs (12) are decomposed into a number of the standard vortex-beams transmitting with different propagation constants. The major perturbation to the ½-charged vortex state contributes to the singly charged vortex [31]. The field distribution looses its initial axial symmetry at the distance much smaller than the beating length $\Lambda$.

The important feature of the nondiffracting vortex-beams is their total angular momentum (TAM) [15, 18, 35, 36]. In the paraxial approximation, we can consider separately the orbital (OAM) and spin (SAM) angular momentum. Berry et. al [14] have shown that the initial beam with the zero OAM and the $1\hbar$ SAM per photon are transformed into the zero SAM and $1\hbar$ OAM in one of the circularly polarized component outside the crystal so that the TAM of the beam is equal to $1/2\hbar$ per photon. The difference between the input and output angular momenta is compensated by the torque excreted on the crystal from the beam. In our case, the TAM of the eigen mode beams in both types of the crystals equals to $(m + 1/2)\hbar$ per photon and is unchanged upon propagation inside the crystal. However, the combined beams consisting of several eigen modes can change periodically the value of their angular momentum.

## IV. Conclusions

We have considered the paraxial propagation of the nondiffracting vortex beams both inside the natural biaxial birefringent medium and the biaxially-induced crystals in vicinity of the optical axis. We have analyzed the solutions to the paraxial wave equation in the form of mode beams bearing optical vortices with different types of topological charges propagating at small angles to the crystal optical axis. All eigen mode beams carry over the total angular momentum that remains unchanged when propagating. The obtained results show that the propagation constants and wave parameters of the vortex-beams in the natural biaxial crystals do not depend on the vortex topological charges defining exclusively with the crystal parameters, the propagation direction and the initial parameters of the beam. In the biaxially-induced crystals the propagation constants are defined by the order of the vortex-beam.

We have revealed a series of new optical effects manifesting themselves in the combined singular beams. For example, the linearly polarized Bessel beam of the zero order at the initial *z=0* plane propagating at the defined angle to the crystal optical axis can be converted either into radially- or azimuthally polarized beam in depending on the orientation angle of the initial linear polarization. Also the converted field can be smoothly transformed into the spirally polarized beams.

We have revealed the steady propagation of the paraxial beams bearing the coupled optical vortices with the fractional topological charges in the circularly polarized components, the total angular momentum of the beam remaining constant along the crystal. The breaking of the vector coupled state results in distortion of the vortex beam with transforming it into a composition of the vortex beams with the standard optical vortices. We have regarded possibilities of the experimental observation of such unusual wave structures. We have shown that the coupled vortices in the beam form the space variant polarization distributions with a linear polarization at each field point and topological index of the centered polarization singularity equal to -1/2.

We have revealed to our great surprise that structure of nondiffracting paraxial mode beams in the biaxially-induced vortex-beams have much to do with that in the conical diffraction process in the natural biaxial crystals. At the same time, there are a number of essential differences. The major of their is that the propagation directions of the conical diffracted mode beams in the natural biaxial crystals are strongly defined by the initial values of the wave parameters altering with the change of the initial beam parameters while the mode beams inside the biaxially-induced crystals do not change their direction along the initial propagation axis irrespective of their wave parameters. Besides, the polarization corrections to the propagation constants depend on the vortex topological charges. As a result, the half-integer vortex-beams in the circularly polarized field components are not stable when propagating along the biaxially-induced crystals.

## V. Acknowledgement

A. Volyar is thankful to R.O. Vlokh for fruitful lengthy discussions about the beam propagation in the biaxial crystals. The paper was partially supported by the grant №0109U002370 of the Ministry of Education, Science, Youth and Sports of Ukraine.

**Figure captures**

Fig.1 Internal conical refraction.

Fig.2 Dispersion curves $\beta(\theta)/k$ and $V(\theta)/k$ in the aragonite crystal with $n_{11} = 1.533$, $n_{22} = 1.686$, $n_{22} = 1.691$, **oa** – optical axis.

Fig.3 (a) The total intensity distribution in the vortex-beam in the state $(1,1,1)$ with $m = 2$, $V = 2400\, m^{-1}$ in the aragonite crystal; (b) The polarization distribution against the background of the intensity distribution in the LHP component.

Fig. 4 (a) Conversion of the radially polarized beam into the zero-order Bessel beam with the uniformly distributed linear polarization and (b) conversion of the linearly zero-order Bessel beam into the spirally polarized beam against a background of the intensity distribution in the RHP component, $V = 5000\, m^{-1}$.

Fig.5 Intensity (a) and polarization (b) distributions for the linearly polarized half-integer vortex-beam at the $z = 0$ plane with $V = 2400\, m^{-1}$.

Fig. 6 Evolution of the space variant linear polarization, $V = 2400\, m^{-1}$.

Fig.7 Evolution of the intensity distribution in the beam with the initial linear polarization and interference patterns of the RHP and LHP components at the quarter beating length $\Delta\beta\, z = \pi/2$, $V = 2400\, m^{-1}$.

Fig. 8 Evolution of the polarization distribution in the uniformly linear polarized beam with the initial uniform linear polarization against the background of the RHP component, $V = 2400\, m^{-1}$.

Fig.9 Sketch of the twisting of the uniaxial crystal (a) and the distribution of the polarization states in the output field $\mathbf{E}_{out}$.

Fig.10 Sketch of the optical axes positions in the twisting uniaxial crystal.

Fig.11 The amplitude $\Psi(R)$ (a) and the intensity (b) distribution in the nondiffracting Airy beam in the biaxially-induced LiNbO$_3$ crystal $U = 3$, $X = x/\rho$, $Y = y/\rho$.

Fig.12 Dispersive curves $\Delta\beta(U)$: $M_z = 0.162\, N \times m$, $n_o = \sqrt{\varepsilon} = 2.3$, $|\pi_{14}| = 8.87 \cdot 10^{-13}\, m^2/N$, $\rho = 0.5\, cm$.

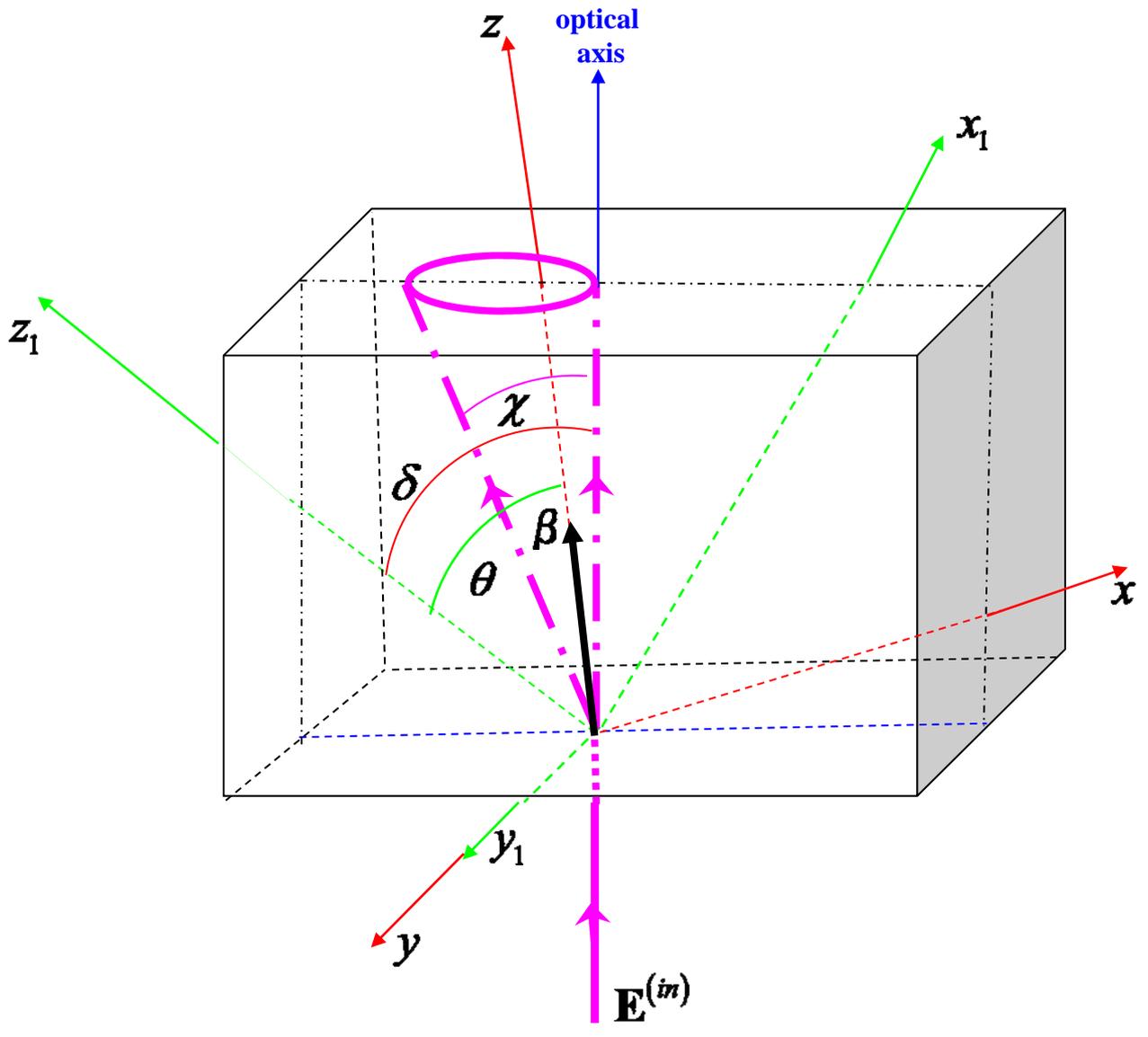

*Fig.1*

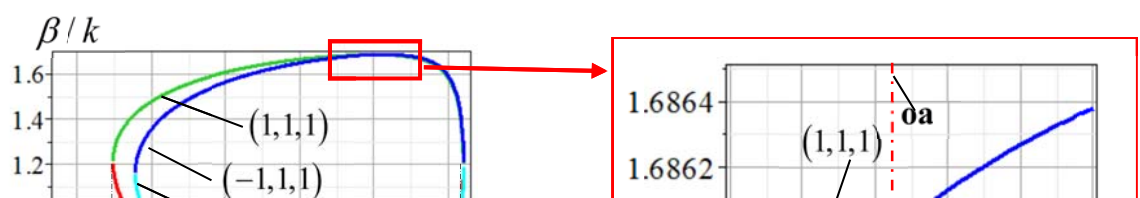

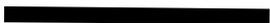
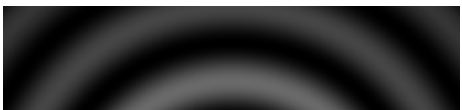
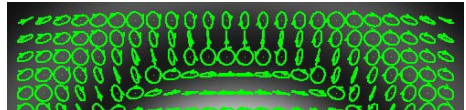

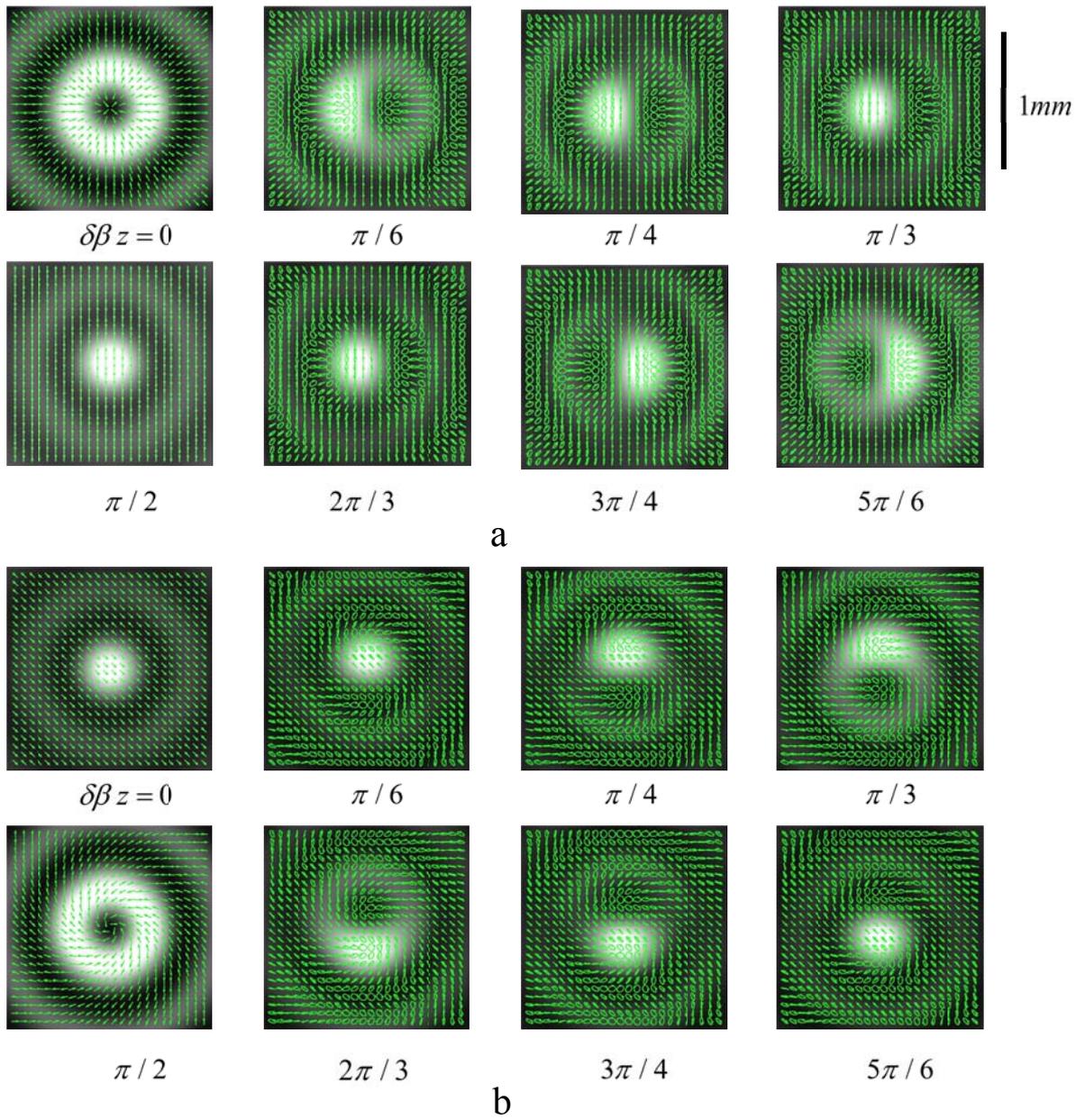

Fig. 4

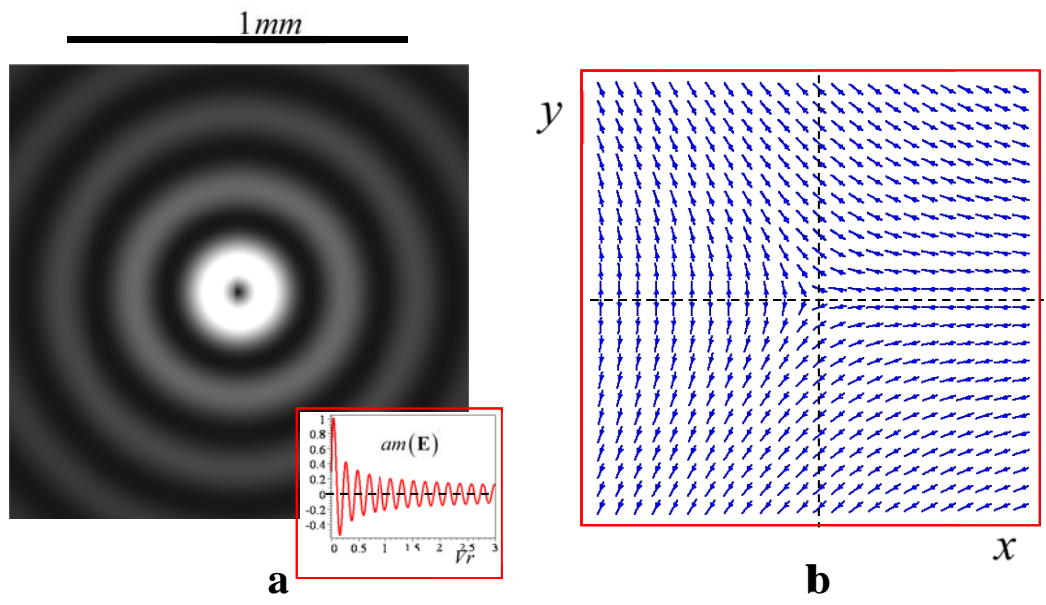

Fig.5

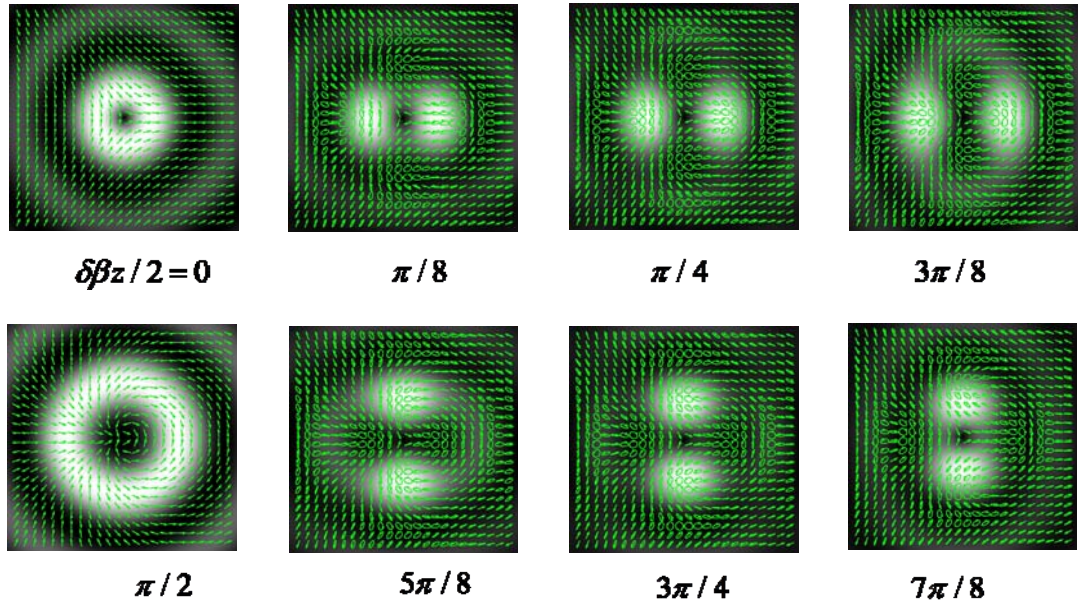

Fig. 6

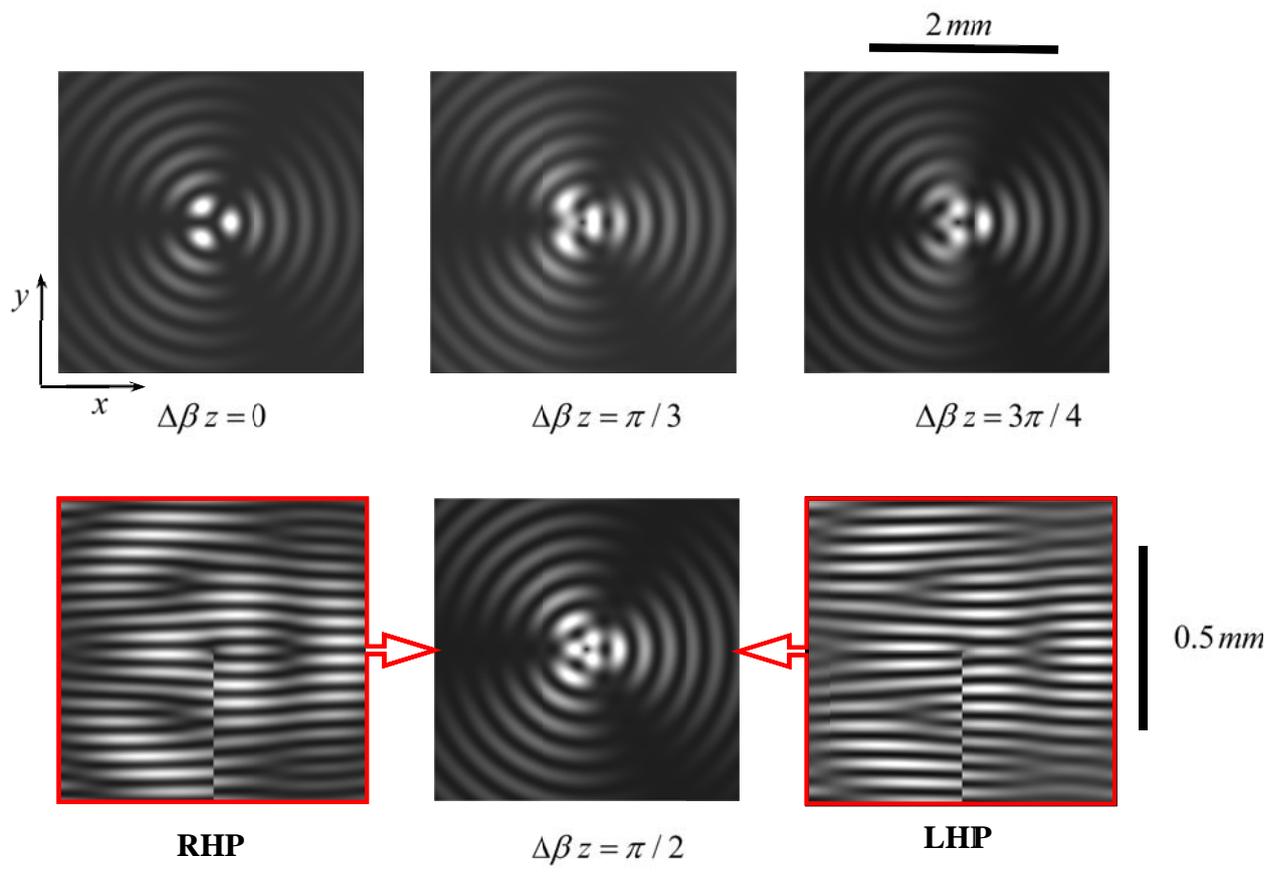

Fig.7

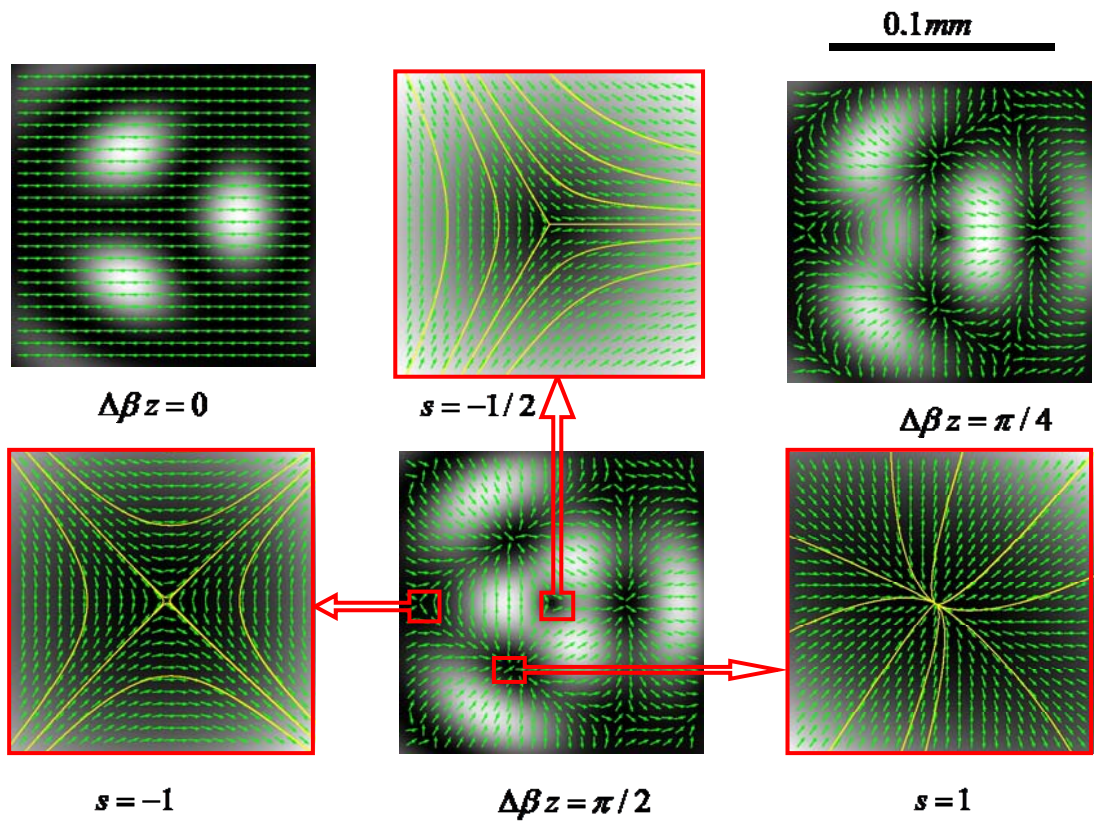

Fig. 8

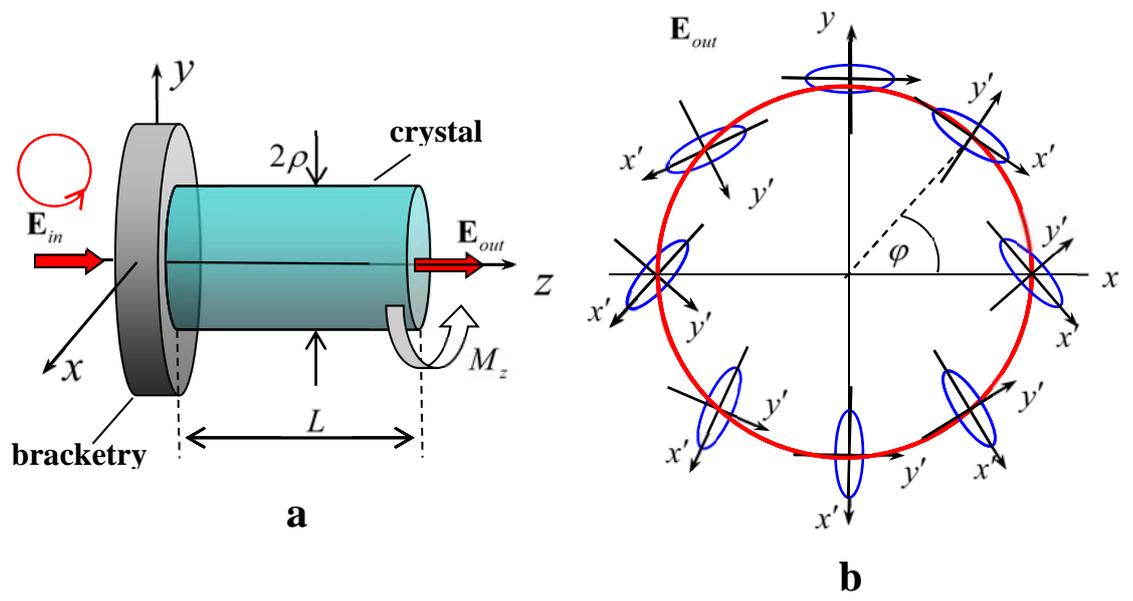

Fig.9

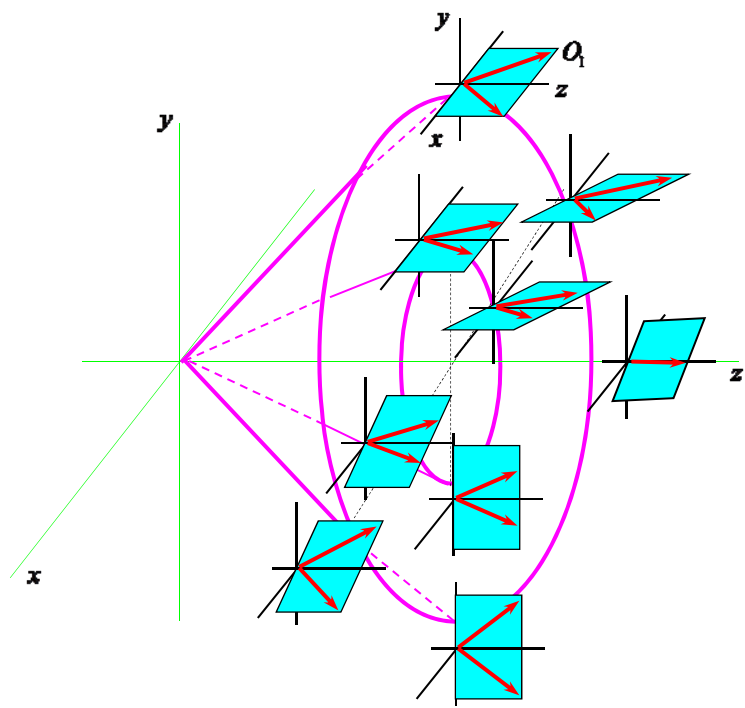

Fig.10

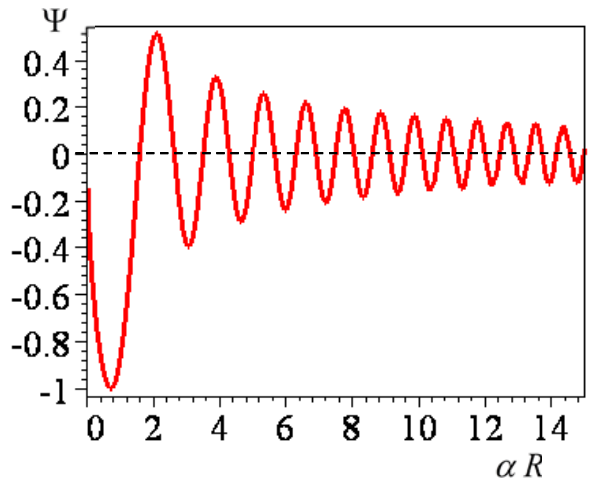 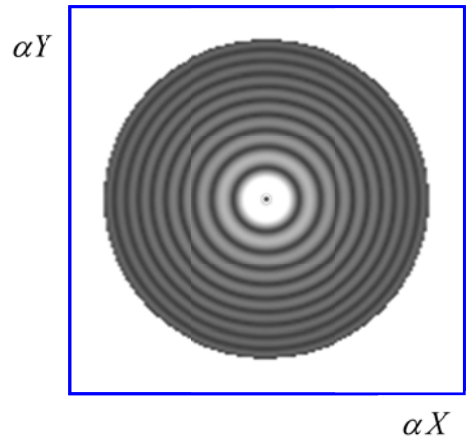

Fig.11

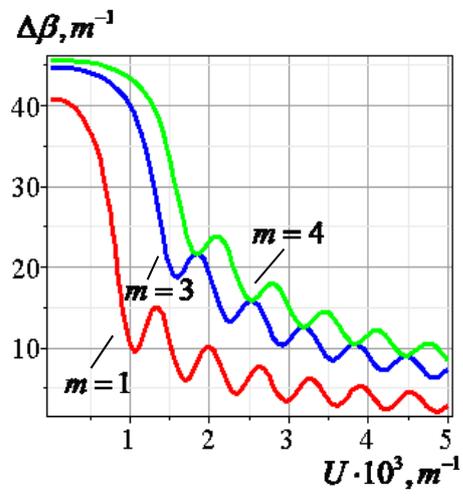

Fig.12